\documentclass[final,3p,times,sort&compress]{elsarticle}

\usepackage{amssymb}
\usepackage{amsthm}

\journal{}

\usepackage{amsfonts}
\usepackage{indentfirst}
\usepackage{graphicx}
\usepackage{slashed}
\usepackage{breqn}
\usepackage[colorlinks = true,
linkcolor = blue,
urlcolor  = blue,
citecolor = blue,
anchorcolor = blue]{hyperref}
\usepackage{flushend}
\usepackage{cuted}
\usepackage{dcolumn}
\usepackage{mathrsfs}
\usepackage{latexsym}

\DeclareMathOperator*{\SumInt}{%
	\mathchoice%
	{\ooalign{$\displaystyle\sum$\cr\hidewidth$\displaystyle\int$\hidewidth\cr}}
	{\ooalign{\raisebox{.14\height}{\scalebox{.7}{$\textstyle\sum$}}\cr\hidewidth$\textstyle\int$\hidewidth\cr}}
	{\ooalign{\raisebox{.2\height}{\scalebox{.6}{$\scriptstyle\sum$}}\cr$\scriptstyle\int$\cr}}
	{\ooalign{\raisebox{.2\height}{\scalebox{.6}{$\scriptstyle\sum$}}\cr$\scriptstyle\int$\cr}}
}
\begin{document}

\begin{frontmatter}



\title{Gravitational chiral anomaly and anomalous transport for fields with spin 3/2}


\author[label1,label2]{G. Yu. Prokhorov}
\ead{prokhorov@theor.jinr.ru}
\author[label1,label2]{O. V. Teryaev}
\ead{teryaev@theor.jinr.ru}
\author[label2,label1,label3]{V. I. Zakharov}
\ead{vzakharov@itep.ru}

\affiliation[label1]{organization={Joint Institute for Nuclear Research},
            addressline={Joliot-Curie 6}, 
            city={Dubna},
            postcode={141980}, 
            country={Russia}}
\affiliation[label2]{organization={Institute of Theoretical and Experimental Physics, NRC Kurchatov Institute},
            addressline={B. Cheremushkinskaya 25}, 
            city={Moscow},
            postcode={117218}, 
            country={Russia}}
\affiliation[label3]{organization={Pacific Quantum Center,
Far Eastern Federal University},
            addressline={10 Ajax Bay, Russky Island}, 
            city={Vladivostok},
            postcode={690950}, 
            country={Russia}}

\begin{abstract}
In a fluid with vorticity and acceleration, an axial current arises in the third order of gradient expansion, called the kinematical vortical effect (KVE). While existing in the absence of gravitational fields, it is nevertheless associated with effects in curved space-time, namely with the gravitational chiral quantum anomaly. In this paper, the KVE transport coefficients were found using the Zubarev quantum-statistical density operator for the Rarita-Schwinger-Adler theory, which includes fields with spins 3/2 and 1/2. A prediction is made about the possible form of the transport coefficients for massless fields with arbitrary spin.
\end{abstract}



%
%
%

\end{frontmatter}


\section{Introduction}
\label{sec intro}

Quantum anomalies play an important role in fundamental physics largely due to the fact that they simultaneously describe processes in different energy regions. Confirmation of such universality has recently been obtained from the physics of quantum fluids - although hydrodynamics, being a gradient expansion, is a kind of low-energy theory, it is to be modified by quantum anomalies \cite{Son:2009tf, Fukushima:2008xe, Sadofyev:2010is, Buzzegoli:2020ycf, Yang:2022ksq, Mitra:2021wjp, Nair:2011mk, Nissinen:2021gke, Chernodub:2021nff}. The corresponding effects can be observed even experimentally - the necessary extreme conditions can be achieved in heavy ion collisions. In particular, the anomalies should contribute to the vorticity-induced polarization \cite{Rogachevsky:2010ys}, studied in various approaches \cite{Becattini:2022zvf}, which is in agreement with the recent data for hyperons with spin 1/2 \cite{STAR:2017ckg}\footnote{Similar experimental data have recently been obtained for hadrons with spin 3/2 \cite{STAR:2020xbm}.}.

It has long been known that accounting for dissipation leads to gradient corrections \cite{Landau1987Fluid}. It has recently been shown that there are also gradient corrections without dissipation. A well-known example is the chiral vortical effect (CVE) \cite{Son:2009tf, Sadofyev:2010is, Yang:2022ksq}, according to which there should be a current directed along the angular velocity
\begin{eqnarray}
j^{\nu}_{A,CVE} = (\sigma_{T} T^2 + \sigma_{\mu} \mu^2) \omega^{\nu} + \mathcal{O}(\partial^3 u) \,,
\label{cve 12}
\end{eqnarray}
where $ T $ is temperature, $ \mu $ is chemical potential, $ \omega^{\mu} ={\frac{1}{2}\varepsilon^{\mu\nu\alpha\beta}
u_{\nu}\partial_{\alpha}u_{\beta}} $ is vorticity, $ \mathcal{O}(\partial^3 u) $ denotes terms cubic in gradients of  4-velocity $ u_{\mu} $ (e.g. $ \omega^ 2\omega^{\nu} $), and $ \sigma_{T} $ and $ \sigma_{\mu} $ are vortical conductivities. For massless Dirac field $ \sigma_{T}=1/6 $, and $ \sigma_{\mu}=1/(2 \pi^2) $. If we now consider the well-known chiral anomaly in the gauge field  $ F_{\mu\nu} $ (we consider the abelian case)
\begin{eqnarray}
\partial_{\mu}j^{\mu}_A = C\, \varepsilon^{\mu\nu\alpha\beta}F_{\mu\nu}F_{\alpha\beta} \,,
\label{anom gauge gen}
\end{eqnarray}
where C is numerical factor (e.g., $ C=-1/(16 \pi^2) $ for the Dirac field), then $ \sigma_{\mu} $ is equal to the anomalous factor
\begin{eqnarray}
\sigma_{\mu} = -8 C \,.
\label{theorem1}
\end{eqnarray}
This statement is not trivial, since the current (\ref{cve 12}) exists even in the absence of a gauge field. However, it was clearly shown in different approaches, e.g. in \cite{Son:2009tf} it was derived from the second law of thermodynamics using the prescription, described in \cite{Landau1987Fluid} (see also the alternative derivation in \cite{Sadofyev:2010is}). Undoubtedly, the most surprising thing is that we have a non-dissipative current in a rotating medium, related to the microscopic topological quantum field properties.

At this stage, a natural question arises - are there hydrodynamic manifestations of other quantum anomalies? For example, there is a gravitational chiral anomaly (also called a mixed axial-gravitational) \cite{Alvarez-Gaume:1984, Duff:1982yw}
\begin{eqnarray}
\nabla_{\mu} j_A^{\mu} = \mathscr{N}{\epsilon}^{\mu\nu\alpha\beta} R_{\mu\nu\lambda\rho}{R_{\alpha\beta}}^{\lambda\rho}\,,
 \label{anom grav gen}
\end{eqnarray}
where $ R_{\mu\nu\lambda\rho} $ is the Riemann tensor, $ {\epsilon}^{\mu\nu\alpha\beta}=\frac{1}{\sqrt{-g}} \varepsilon^{\mu\nu\alpha\beta}$ is the Levi-Civita symbol in curved space, $ \nabla_{\mu} $ is the covariant derivative, and $ \mathscr{N} $ is a numerical factor. In particular, for the Dirac fields $ \mathscr{N} = 1/(384 \pi^2) $.

The answer to this question turned out to be rather nontrivial and was obtained in \cite{Landsteiner:2011cp, Jensen:2012kj, Stone:2018zel}. 
The condition of the translational
invariance of the Euclidean vacuum leads to the relation $ \sigma_{T} = 64 \pi^2 \mathcal{N}$.
This formula has been explicitly verified for fields with spin 1/2, where $ \sigma_{T} = 64 \pi^2 \mathcal{N} = 1/6 $, and also for spin 1 field \cite{Prokhorov:2020npf}.

However, there are problems when considering higher spins.
Namely, in the recently proposed Rarita-Schwinger-Adler theory, describing fields with spins 3/2 and 1/2 \cite{Adler:2017shl}, the chiral anomaly has the form
\begin{eqnarray}
\nabla_{\mu} j^{\mu}_{A,RSA} =-\frac{5}{16\pi^2 \sqrt{-g}}\varepsilon^{\mu\nu\alpha\beta}F_{\mu\nu}F_{\alpha\beta} -\frac{19}{384 \pi^2\sqrt{-g}}\varepsilon^{\mu\nu\rho\sigma}
R_{\mu\nu\kappa\lambda}{R_{\rho\sigma}}^{\kappa\lambda}\,,\,\,\,
\label{anom rsa}
\end{eqnarray}
where the gauge part was found in \cite{Adler:2017shl} (see also \cite{Adler:2019zxx, Prokhorov:2022rna}), and the gravitational part  was found in \cite{Prokhorov:2022rna}. At the same time, the CVE current in the RSA theory has the form \cite{Prokhorov:2021bbv}
\begin{eqnarray}
j^{\nu}_{A,CVE} = \left(\frac{5}{6} T^2 + \frac{5}{2 \pi^2} \mu^2 \right) \omega^{\nu} + \mathcal{O}(\partial^3 u)\,.\,\,\,
\label{cve 32}
\end{eqnarray}
Comparing (\ref{anom rsa}) and (\ref{cve 32}), one can immediately notice that the anomaly-conductivity correspondence (\ref{theorem1}) for the gauge anomaly is directly confirmed. However, there is a problem when compared with the gravitational anomaly, since $ \sigma_T$ differs from the factor $64 \pi^2 \mathcal{N} $ both in absolute value and in sign. The large negative number that is due to the rapid cubic growth of the anomalous factor with spin (see Section \ref{sec disc}) is not visible.

At the same time, the detection of the number ``$ -19 $''  from (\ref{anom rsa}) in hydrodynamics would become a clear marker of a gravitational anomaly. With this goal in mind, we will consider a recent result \cite{Prokhorov:2022udo}. It was shown that the current arising in the third order of the gradient expansion
\begin{eqnarray}
j^{\mu}_{A,KVE} &=& \lambda_1 (\omega_{\nu}\omega^{\nu})\omega^{\mu} + \lambda_2 (a_{\nu}a^{\nu})\omega^{\mu}\,,
 \label{kve gen}
\end{eqnarray}
where $ a_{\mu}=u^{\nu}\partial_{\nu} u_{\mu} $ is acceleration, and $ \lambda_1 $ and $ \lambda_2 $ are corresponding conductivities, satisfies the key relation
\begin{eqnarray}
\lambda_{1}-\lambda_{2}&=& 32 \mathscr{N}\,.
\label{theorem2}
\end{eqnarray}
In this case, the factor from the gravitational chiral anomaly is split into two transport coefficients, in contrast to the case with the gauge anomaly  (\ref{theorem1}). The current (\ref{kve gen}) was called the kinematical vortical effect (KVE), since explicitly it depends only on kinematical quantities.

In \cite{Prokhorov:2022udo} the relation (\ref{theorem2}) was derived in the general spin-independent case and then tested for spin 1/2. Our goal is to derive explicitly the KVE for spin 3/2 and check its relation to the anomaly using (\ref{theorem2}). However, first, let's recall how (\ref{theorem2}) was obtained.

We use the notation and conventions $ \delta_{\mu\nu} = {\text{diag}(1,1,1,1)} $, $ \eta_{\mu\nu} = {\text{diag}(1,-1,-1,-1)} $, $\varepsilon^{0123}=1$, and the system of units $ e=\hbar=c=k_B=1 $.

\section{Anomalous transport associated with the gravitational chiral anomaly}
\label{sec gen deriv}

In this section, we describe the key points of the derivation of (\ref{theorem2}) while the details are given in \cite{Prokhorov:2022udo}. Consider an uncharged fluid of massless fermions with an arbitrary spin in an external gravitational field, characterized by the inverse proper temperature vector $ \beta_{\mu}=u_{\mu}/T $, where $ T $ is the proper temperature and $ u_{\mu} $ is the 4-velocity of the medium. Hydrodynamics is constructed as a gradient expansion, and the expansion for the axial current in the third order is
\begin{eqnarray}
j_{\mu}^{A(3)} = \xi_{1}(T) w^2 w_{\mu} + \xi_{2}(T) \alpha^2 w_{\mu}  + \xi_{3}(T) (\alpha w) w_{\mu}  + \xi_{4}(T) A_{\mu\nu}w^{\nu}+ \xi_{5}(T) B_{\mu\nu}a^{\nu}\,,
\label{current decomp}
\end{eqnarray}
where, in particular, $ w^2=w_{\nu} w^{\nu} $ and $ (\alpha w)=\alpha_{\nu} w^{\nu} $, and the notation is used
\begin{eqnarray}
\varpi_{\mu\nu} = -\frac{1}{2} (\nabla_{\mu} \beta_{\nu} - \nabla_{\nu} \beta_{\mu}) \,, \,\,\,
\,\alpha_{\mu} = \varpi_{\mu\nu} u^{\nu} \,, \,\,\,
w_{\mu} = -\frac{1}{2} {\epsilon}_{\mu\nu\alpha\beta} u^{\nu} \varpi^{\alpha \beta}\,, \,\,\, A_{\mu\nu} = u^{\alpha}u^{\beta} R_{\alpha\mu\beta\nu}\,, \,\,\,
B_{\mu\nu} = \frac{1}{2} {\epsilon}_{\alpha\mu\eta\rho} u^{\alpha}u^{\beta} {R_{\beta\nu}}^{\eta\rho}\,.
\label{tensors}
\end{eqnarray}
Here $ \varpi_{\mu\nu} $ is the thermal vorticity tensor, $ w_{\mu}=\omega^{\mu}/T $ and $ \alpha_{\mu} $ are the ``thermal'' vorticity pseudovector and acceleration, and 
$ \omega^{\mu} ={\frac{1}{2}\epsilon^{\mu\nu\alpha\beta} u_{\nu}\nabla_{\alpha}u_{\beta}} $ is the usual vorticity pseudovector. The expansion (\ref{current decomp}) is valid for the case of global thermodynamic equilibrium in the beta-frame \cite{Becattini:2014yxa}, for which $ \beta_{\mu} $ coincides with the Killing vector
\begin{eqnarray}
\nabla_{\mu} \beta_{\nu} +
\nabla_{\nu} \beta_{\mu} = 0\,.
\label{killing}
\end{eqnarray}
In particular, in the state of global equilibrium $ \alpha^{\mu}=a^{\mu}/T $, where $ a_{\mu}=u^{\nu}\nabla_{\nu} u_{\mu} $ is the usual acceleration. We also use the fact that the gravitational field is external and impose an additional condition $ R_{\mu\nu}=0 $.

The unknown coefficients $ \xi_n (T) $ can be obtained from the equality of the divergence of the current (\ref{current decomp}) to the anomaly (\ref{anom grav gen}) 
\begin{eqnarray}
\nabla^{\mu}\Big(\xi_{1}(T) w^2 w_{\mu} + \xi_{2}(T) \alpha^2 w_{\mu}  + \xi_{3}(T) (\alpha w) w_{\mu} + \xi_{4}(T) A_{\mu\nu}w^{\nu}+ \xi_{5}(T) B_{\mu\nu}a^{\nu}\Big)
= 32 \mathcal{N} A_{\mu\nu} B^{\mu\nu}\,.\quad
\label{condition}
\end{eqnarray}
This approach is a generalization of \cite{Landau1987Fluid} and \cite{Son:2009tf} to the case of curved space-time and gravitational chiral anomaly. However, in contrast to \cite{Landau1987Fluid, Son:2009tf}, we also use the global equilibrium condition, and do not consider the entropy current \cite{Buzzegoli:2020ycf, Yang:2022ksq}. In fact, this simplifies the derivation, allowing us to consider a nontrivial case with gravity.

Equation (\ref{condition}) splits into independent terms, which give a system of the first order differential equations for the coefficients $ \xi_n (T) $. Solving it, we will obtain that $ \xi_n (T) = \lambda_n T^3 $ and $ \lambda_1 $ and $ \lambda_2 $ should satisfy (\ref{theorem2}) (while $ \xi_3 (T) =0 $). The current (\ref{current decomp}) in the flat space limit transforms into (\ref{kve gen}).

In \cite{Prokhorov:2022udo} the relation obtained was verified for massless Dirac fields. From the point of view of various approaches \cite{Prokhorov:2018bql, Ambrus:2021eod, Ambrus:2019ayb, Palermo:2021hlf} (see also \cite{Vilenkin:1979ui, Vilenkin:1980zv}), using the microscopic theory of Dirac field, the following formula for the axial current was obtained
\begin{eqnarray}
j^{\mu}_{A,\text{Dirac}}&=&\left(\frac{T^2}{6}+\frac{\mu^2}{2\pi^2}-\frac{\omega^2}{24 \pi^2} -\frac{a^2}{8 \pi^2} \right)
\omega^{\mu}\,.
\label{kve 12}
\end{eqnarray}
Comparing (\ref{kve 12}) with the gravitational anomaly for the Dirac field $ \mathcal{N}=1/(384 \pi^2) $, we see
\begin{eqnarray}
\lambda_{1}-\lambda_{2}=-\frac{1}{24 \pi^2} + \frac{1}{8 \pi^2} =\frac{32}{384 \pi^2} \,.
\label{check 12}
\end{eqnarray}
Thus, (\ref{theorem2}) is satisfied. However, isn't (\ref{check 12}) just a coincidence? Is it possible to obtain additional confirmation for (\ref{theorem2})? To answer these questions, in the next sections we will consider a much more complicated system.

\section{Rarita-Schwinger-Adler model for fields with spin 3/2}
\label{sec rsa}

The Rarita-Schwinger field theory is one of the most common approaches to describe fields with spin 3/2. But this theory, as it was shown a long time ago, has a number of problems if we use it outside of supergravity. Recently, in \cite{Adler:2017shl}, an extension, the Rarita-Schwinger-Adler (RSA) theory, was developed, which made it possible, at least, to construct a perturbation theory in an external field and to find the gauge chiral anomaly by the well-known shift method. The action in this theory (in flat space-time) has the form
\begin{eqnarray}
S = \int d^4 x\,\Big( -\varepsilon^{\lambda \rho \mu \nu} \bar{\psi}_{\lambda}\gamma_5 \gamma_{\mu} \partial_{\nu} \psi_{\rho}+i \bar{\lambda} \gamma^{\mu}\partial_{\mu} \lambda  - i m \bar{\lambda} \gamma^{\mu}\psi_{\mu}+i m \bar{\psi}_{\mu} \gamma^{\mu}\lambda \Big)\,,
\label{action}
\end{eqnarray}
where $ \psi_{\mu} $ is the Rarity-Schwinger field, $ \lambda $ is the additional Dirac field, $ m $ is the interaction constant 
. 
Following \cite{Adler:2017shl}, 
we will consider the limit $ m \to \infty $. The stress-energy tensor can be constructed by varying the metric
\begin{eqnarray}
T^{\mu\nu}&=& \frac{1}{2} \varepsilon^{\lambda\nu\beta\rho}\bar{\psi}_{\lambda}\gamma_5 \gamma^{\mu}  \partial_{\beta}\psi_{\rho} +\frac{1}{8}\partial_{\eta}\Big(
\varepsilon^{\lambda\alpha\nu\rho}\bar{\psi}_{\lambda}\gamma_5\gamma_{\alpha} [\gamma^{\eta},\gamma^{\mu}] \psi_{\rho}\Big) +\frac{i}{4}\Big(
\bar{\lambda}\gamma^{\nu}\partial^{\mu}\lambda-\partial^{\mu}\bar{\lambda}\gamma^{\nu}\lambda
\Big) \nonumber \\&& +\frac{i }{2}m\Big(
\bar{\psi}^{\mu}\gamma^{\nu}\lambda-\bar{\lambda}\gamma^{\nu}\psi^{\mu} \Big) + (\mu \leftrightarrow \nu)\,,
\label{ten en mom}
\end{eqnarray}
while the axial current can be obtained using Noether's theorem, for the $ U(1)_A $ transformation
\begin{eqnarray}
j^{\mu}_A =  -i \varepsilon^{\lambda\rho\nu\mu}\bar{\psi}_{\lambda}\gamma_{\nu}\psi_{\rho} +\bar{\lambda}\gamma_{\mu}\gamma_5 \lambda\,,
\label{cur}
\end{eqnarray}
and the  chiral quantum anomaly has the form (\ref{anom rsa}). 

\section{KVE in the Rarita-Schwinger-Adler model: the result of the density operator}
\label{sec-calc}

Let us calculate the coefficients $ \lambda_1 $ and $ \lambda_2 $ in (\ref{kve gen}) from the microscopic theory (\ref{action}). To do this, we use the Zubarev quantum statistical density operator \cite{Buzzegoli:2017cqy, Zubarev:1979, Buzzegoli:2020ycf}
\begin{eqnarray}
\hat{\rho}=\frac{1}{Z}\exp\bigg\{-\beta_{\mu}(x)\hat{P}^{\mu}
+\frac{1}{2}\varpi_{\mu\nu}\hat{J}^{\mu\nu}_x
\bigg\} \,,
\label{rho}
\end{eqnarray}
where $ \hat{P}^{\mu} $ is the 4-momentum operator, and $ \hat{J}^{\mu\nu}_x $ are the generators of the Lorentz transformations, shifted by $ x_{\mu} $.  The generators $ \hat{J}^{\mu\nu}_x $ are expressed in terms of the stress-energy tensor
\begin{eqnarray}
\hat{J}^{\mu\nu}_x=
\int d\Sigma_{\lambda}\bigg[y^{\mu}\hat{T}^{\lambda\nu}_x(y)-y^{\nu}\hat{T}_x^{\lambda\mu}(y)\bigg] \,,
\label{gen J}
\end{eqnarray}
where $ d\Sigma_{\lambda}$ is an element of an arbitrary 3-D spacelike hypersurface. Operator $ \hat{J}^{\mu\nu} $ can be decomposed into boost $ \hat{K}^{\mu} $ and orbital angular momentum $ \hat{J}^{\mu} $
\begin{eqnarray}
\hat{J}^{\mu\nu}=u^{\mu}\hat{K}^{\nu}-u^{\nu}\hat{K}^{\mu}-
\epsilon^{\mu\nu\rho\sigma}u_{\rho}\hat{J}_{\sigma}\,,
\label{dec J}
\end{eqnarray}
The operator in the form (\ref{rho}) describes a medium in the state of global thermodynamic equilibrium (\ref{killing}) (in the limit of flat space), which, in particular, allows us to choose $ d\Sigma_{\lambda} = \eta_{0\lambda} d^3 y$. Using (\ref{rho}), different effects were derived \cite{Buzzegoli:2017cqy, Palermo:2021hlf, Prokhorov:2019cik, Prokhorov:2019yft}, including CVE and KVE for Dirac fields \cite{Buzzegoli:2017cqy, Prokhorov:2018bql}, and CVE in RSA theory \cite{Prokhorov:2021bbv}.

The effects associated with the thermal vorticity tensor can be calculated using perturbation theory and finite-temperature field theory. The average of the local operator $ \hat{O}(x) $ is determined by the perturbative series
\begin{eqnarray}
\langle\hat{O}(x)\rangle &=&
\langle\hat{O}(0)\rangle_{\beta(x)}+
\sum_{N=1}^{\infty}\frac{ \varpi^N }{2^N|\beta|^N N!}\int_{0}^{|\beta|}d\tau_1... d\tau_N
 \langle T_{\tau}
\hat{J}_{-i\tau_1 u}...
\hat{J}_{-i\tau_N u}
\hat{O}(0)
\rangle_{\beta(x),c}
\,,
\label{pert O}
\end{eqnarray}
where  $ {|\beta|=\sqrt{|\beta_{\nu}\beta^{\nu}|}= 1/T} $,  $ \langle .. \rangle_{\beta(x),c} $ means the connected part of the grand-canonical ensemble average, $ T_{\tau} $ means ordering by imaginary time $ \tau = i\, t $, the generators are shifted by the vector $ - i\tau_n u^{\mu}$ and each of the operators $ \hat{J} $ is contracted with one of the thermal vorticity tensors: $ \varpi_{\mu\nu}\hat{J}^{\mu\nu} $. It is convenient to introduce the following notation
\begin{eqnarray}
&&X_{\mu}=(\tau_x,-{\bf x})\,,\,\,
\gamma_{\mu}=i^{\delta_{0\mu}-1}\tilde{\gamma}_{\mu}\,,\,\,
\tilde{\gamma}_5=\gamma_5=i\gamma^0\gamma^1\gamma^2\gamma^3\,,\,\, {\partial}_{\mu}=i^{\delta_{0\mu}}\tilde{\partial}_{\mu}\,,\,\,
\tilde{\partial}^{X}_{\mu} = \left(\frac{\partial}{\partial \tau_x},\frac{\partial}{\partial \bold{x}} \right) \,,\,\, \psi_{\mu}=i^{\delta_{0\mu}}\tilde{\psi}_{\mu}\,,
\nonumber \\
&&P_{\mu}=(p_n,-{\bf p})\, , \,\,
p_n=(2n+1) \pi T \quad (n=0,\pm 1,...) \,,\,\, \SumInt_{P}= T \sum_{p_n}\int\frac{d^3p}{(2\pi)^3}\,,\,\,
\slashed{P}=P_{\mu}\tilde{\gamma}_{\mu}\,,\,\, P^2=P_{\mu}P_{\mu}\,,
\label{not}
\end{eqnarray}
where $ p_n $ are the fermionic Matsubara frequencies. Euclidean gamma matrices $ \tilde{\gamma}_{\mu} $ satisfy the relation $ \{\tilde{\gamma}_{\mu},
\tilde{\gamma}_{\nu}\}=2\delta_{\mu\nu} $.

Comparing (\ref{pert O}) (in the third order) and (\ref{kve gen}), moving into the rest frame $ u_{\mu}={(1,0,0,0)} $, and choosing appropriately $ w_{\mu} $ and $ \alpha_{\mu} $ (e.g., $ \alpha_{\mu}=0,\, \omega_{\mu}={(0,0,0,\omega_3)} $ for $ \lambda_{1} $), it is easy to show that \cite{Prokhorov:2018bql}
\begin{eqnarray} \label{lambda KJ}
\lambda_1 &=& -\frac{1}{6}
\int_0^{|\beta|}[d\tau]
\langle T_{\tau} \hat{J}^{3}_{-i\tau_{x}}\hat{J}^{3}_{-i\tau_{y}}\hat{J}^{3}_{-i\tau_{z}}\hat{j}_A^{3}(0)\rangle_{T,c} \,, \\ \nonumber 
\lambda_2 &=&
-\frac{1}{6} \int_0^{|\beta|} [d\tau]
\bigg\{
 \langle T_{\tau}(\hat{K}^{1}_{-i\tau_{x} }\hat{J}^{3}_{-i\tau_{y}} + \hat{J}^{3}_{-i\tau_{x} }\hat{K}^{1}_{-i\tau_{y}})\hat{K}^{1}_{-i\tau_{z} u}\hat{j}^{3}_A(0)\rangle_{T,c} + \langle T_{\tau} \hat{K}^{1}_{-i\tau_{x} }\hat{K}^{1}_{-i\tau_{y} }\hat{J}^{3}_{-i\tau_{z} }\hat{j}^{3}_A(0)\rangle_{T,c}
\bigg\}\,,
\end{eqnarray}
where $ [d \tau] = d \tau_x d \tau_y d \tau_z$ (we've changed the notation slightly to show the transition to the rest frame) \footnote{The coefficient $ \xi_3=\lambda_3 T^3 =0$ can be taken to be zero in advance.
}. Let us introduce the following typical correlator (the temperature factor is introduced to match the accepted notation)
\begin{eqnarray}
C^{\alpha_1\alpha_2|\alpha_3\alpha_4|\alpha_5\alpha_6|\lambda|ijk} =
T^3 \int [d\tau] d^3x\, d^3y\, d^3z\, x^i y^j z^k
\langle
T_{\tau}
\hat{T}^{\alpha_1\alpha_2}(-i\tau_x,\bold{x}) 
\hat{T}^{\alpha_3\alpha_4}(-i\tau_y,\bold{y})
\hat{T}^{\alpha_5\alpha_6}(-i\tau_z,\bold{z}) \hat{j}_5^{\lambda}(0)\rangle_{T, c} \,.\quad \quad
\label{C}
\end{eqnarray}
From (\ref{lambda KJ}), taking into account (\ref{gen J}), we will obtain
\begin{eqnarray} 
\lambda_1 &=& -\frac{1}{6 T^3}\bigg(
C^{02|02|02|3|111}+C^{02|01|01|3|122}  +C^{01|02|01|3|212} +C^{01|01|02|3|221}\nonumber \\
&& -C^{01|01|01|3|222}  -C^{01|02|02|3|211}-C^{02|01|02|3|121}-C^{02|02|01|3|112}\bigg) \,, \nonumber \\
\lambda_2 &=& -\frac{1}{6 T^3}\bigg(
C^{02|00|00|3|111}+C^{00|02|00|3|111} +C^{00|00|02|3|111}-C^{01|00|00|3|211}\nonumber \\
&& -C^{00|01|00|3|121}-C^{00|00|01|3|112}\bigg)  \,.
\label{lambda C}
\end{eqnarray}
Let's describe the main stages of calculation of the four-point function (\ref{C}). Combining all fields into a vector with 5 indices $ \Psi_{I}=\{\tilde{\psi}_{\mu},\lambda\} $, where ${ (I=0 ... 4 )}$, and also splitting the point in the local operator, we will obtain
\begin{eqnarray} \nonumber
\hat{T}^{\mu\nu}(X)&=&\lim_{X_1,X_2\to X} \mathcal{D}^{\mu\nu}_{a b (IJ)}(\tilde{\partial}_{X_1},\tilde{\partial}_{X_2})\bar{\Psi}_{aI}(X_1)\Psi_{bJ}(X_2) \,, \nonumber \\
\hat{j}_A^{\mu}(X)&=&\lim_{X_1,X_2\to X} \mathcal{J}^{\mu}_{a b (IJ)}\bar{\Psi}_{aI}(X_1)\Psi_{bJ}(X_2) \,,
\label{split}
\end{eqnarray}
where $ a $ and $ b $ are bispinor indices and the summation over repeated indices is assumed. The operators $ \mathcal{D} $ and $ \mathcal{J} $ are determined from (\ref{ten en mom}), (\ref{cur}) and (\ref{not})
\begin{eqnarray}
\mathcal{D}^{\mu\nu}_{(ij)}&=& -\frac{1}{2} (-i)^{\delta_{0\mu}+\delta_{0\nu}}\varepsilon^{i j \nu\beta}\left(\gamma_5 \tilde{\gamma}_{\mu} \tilde{\partial}_{\beta}^{X_2}-\frac{1}{4}\gamma_5 \tilde{\gamma}_{\beta}[\tilde{\gamma}_{\vartheta},\tilde{\gamma}_{\mu}] \left(\tilde{\partial}^{X_1}_{\vartheta}+\tilde{\partial}^{X_2}_{\vartheta}\right)\right)+(\mu\leftrightarrow \nu)\,,
\nonumber \\
\mathcal{D}^{\mu\nu}_{(i4)}&=& -\mathcal{D}^{\mu\nu}_{(4i)} =
\frac{m}{2} (-i)^{\delta_{0\mu}+\delta_{0\nu}}
\tilde{\gamma}_{\nu}\delta_{\mu i}+(\mu\leftrightarrow \nu)\,,
\nonumber \\
\mathcal{D}^{\mu\nu}_{(44)}&=&
\frac{1}{4} (-i)^{\delta_{0\mu}+\delta_{0\nu}}
\tilde{\gamma}_{\mu}(\tilde{\partial}_{X_2}-\tilde{\partial}_{X_1})_{\nu}+(\mu\leftrightarrow \nu)\,,
\nonumber \\
\mathcal{J}^{\mu}_{(ij)}&=& i^{1-\delta_{0\mu}}\varepsilon^{i j \mu \nu}
\tilde{\gamma}_{\nu}\,,
\nonumber \\
\mathcal{J}^{\mu}_{(i4)}&=& \mathcal{J}^{\mu}_{(4i)} = 0\,,
\nonumber \\
\mathcal{J}^{\mu}_{(44)}&=& i^{1-\delta_{0\lambda}}
\tilde{\gamma}_{\lambda} \gamma_{5}
\,,
\label{split 1}
\end{eqnarray}
where ${0 \leq (i, j)<4 }$. We will also need the finite-temperature Green's functions, which were found in \cite{Prokhorov:2021bbv} (in complete analogy with \cite{Adler:2017shl}) from the functional integral (for simplicity, we consider the case $ \mu=\mu_A=0 $)
\begin{eqnarray}
&&\langle T_{\tau}\tilde{\psi}_{a \mu}(X_1)\tilde{\bar{\psi}}_{b\nu}(X_2)\rangle_{T} = \SumInt_{P}e^{iP_{\alpha}(X_1-X_2)^{\alpha}}\frac{i}{2 P^2} \left(\tilde{\gamma}_{\nu} \slashed{P} \tilde{\gamma}_{\mu} +2\left[\frac{1}{m^2}-\frac{2}{P^2}\right]
P_{\mu}P_{\nu}\slashed{P}\right)_{ab}\,,\nonumber \\
&&\langle T_{\tau}\lambda_a(X_1)\tilde{\bar{\psi}}_{b\mu}(X_2)\rangle_{T} = \SumInt_{P}e^{iP_{\alpha}(X_1-X_2)^{\alpha}}\frac{P_{\mu}\slashed{P}_{ab}}{m  P^2 }\,,\nonumber \\
&&\langle T_{\tau}\tilde{\psi}_{a\mu}(X_1)\bar{\lambda}_b(X_2)\rangle_{T} = \SumInt_{P}e^{iP_{\alpha}(X_1-X_2)^{\alpha}}\frac{-P_{\mu}\slashed{P}_{ab}}{m  P^2 }\,,\nonumber \\
&&\langle T_{\tau}\lambda_a(X_1)\bar{\lambda}_b(X_2)\rangle_{T} = 0\,,
\label{prop}
\end{eqnarray}
where only the exponent contains the summation with the non-Euclidean metric $ (P_{\mu} X^{\mu} = p_n \tau_x - \bold{p}\bold{x})$ \cite{Laine:2016hma}. Substituting (\ref{split}) into (\ref{C}), we obtain
\begin{eqnarray}
C^{\alpha_1\alpha_2|\alpha_3\alpha_4|\alpha_5\alpha_6|\lambda|ijk} &=& - T^3 \int [d\tau]d^3 x d^3 y d^3 z \, x^{i} y^{j} z^k
 \lim_{X,Y,Z,F} \mathcal{D}^{\alpha_1\alpha_2}_{a_1 a_2(I_1 I_2)}(\tilde{\partial}_{X_1},\tilde{\partial}_{X_2})
\mathcal{D}^{\alpha_3\alpha_4}_{a_3 a_4(I_3 I_4)}(\tilde{\partial}_{Y_1},\tilde{\partial}_{Y_2}) \mathcal{D}^{\alpha_5\alpha_6}_{a_5 a_6(I_5 I_6)}(\tilde{\partial}_{Z_1},\tilde{\partial}_{Z_2})\nonumber \\
&& \times\mathcal{J}^{\lambda}_{a_7 a_8(I_7 I_8)}
\langle T_{\tau} \overline{\Psi}_{a_1 I_1}(X_1)\Psi_{a_2 I_2}(X_2 ) \overline{\Psi}_{a_3 I_3}(Y_1)\Psi_{a_4 I_4}(Y_2) \overline{\Psi}_{a_5 I_5}(Z_1)\Psi_{a_6 I_6}(Z_2)\overline{\Psi}_{a_7 I_7}(F_1) \Psi_{a_8 I_8}(F_2)\rangle_{T,c}\,,\qquad
\label{C split}
\end{eqnarray}
where the limit $ {(X_1,X_2 \to X)}$,  $ {(Y_1,Y_2 \to Y)}$, $ {(Z_1,Z_2 \to Z)}$ and $ {(F_1,F_2 \to 0)}$ is taken. Using Wick's theorem
\begin{eqnarray}
\langle \overline{\Psi}(X)\Psi(X)\overline{\Psi}(Y)\Psi(Y)\overline{\Psi}(Z)\Psi(Z)\overline{\Psi}(F)\Psi(F) \rangle_c = -\langle \Psi(Y)\overline{\Psi}(X)\rangle\langle\Psi(X)\overline{\Psi}(F)\rangle
\langle\Psi(Z)\overline{\Psi}(Y) \rangle
\langle \Psi(F)\overline{\Psi}(Z)\rangle
  +\, \text{(5 terms)}\,,\quad
\label{wick}
\end{eqnarray}
we will obtain six connected terms (as usual, the sign is determined by the number of permutations). Then (\ref{C split}) will take the form of the product of propagators and vertices. A direct enumeration of all the values of the indices $ I_n $ in (\ref{C split}) would result in $6\times 5^8 $ terms. However, many of them are zero, because:
\begin{itemize}
\item some vertices are zero, like $ \mathcal{J}^{\mu}_{(i4)}=\mathcal{J}^{\mu}_{(4i)}=0 $, and also $ \mathcal{D}^{\mu\nu}_{(ij)} = 0$ if $ i=j\neq 4 $;
\item some terms are zero because they include propagator $ \langle \lambda \bar{\lambda} \rangle =0  $;
\item some terms are zero in the limit $ m \to \infty $ if the total number of propagators $ \langle \lambda \bar{\psi} \rangle $ and $ \langle \psi \bar{\lambda} \rangle $ is greater than the total number of vertices $ \mathcal{D}_{(i4)}$ and $ \mathcal{D}_{(4i)}$. We can also drop the terms $ 1/m^2 $ from the first propagator in (\ref{prop}).
\end{itemize}
Because of this, for each $ C $-correlator from $ \lambda_1 $, 94752 terms remain, and each of the $ C $-correlators in $ \lambda_2 $ contains 31152 terms. We used Wolfram Mathematica to calculate them.

To illustrate the next steps of the calculation, let's consider the contribution to (\ref{C split}) from the first term in (\ref{wick})
\begin{eqnarray}  
C_{\text{Wick1}}^{\alpha_1\alpha_2|\alpha_3\alpha_4|\alpha_5\alpha_6|\lambda|ijk} &=&- T^3 \int [d\tau]d^3 x d^3 y d^3 z\, x^i y^j z^k \SumInt_{{P_1,P_2,P_3,P_4}}
e^{i P_1(Y-X)+i P_2 X+i P_3 (Z-Y)-i P_4 Z} \Delta(P_1)\Delta(P_2)\Delta(P_3)\Delta(P_4) \nonumber \\
&& \times \underset{I,a}{\text{tr}} \Big\{ \mathcal{D}^{\alpha_1\alpha_2}(-i P_1, i P_2) G(P_2) \mathcal{J}^{\lambda} G(P_4)
\mathcal{D}^{\alpha_5\alpha_6}(-i P_4, i P_3)  G(P_3)\mathcal{D}^{\alpha_3\alpha_4}(-i P_3, i P_1) G(P_1)\Big\} \,,
\label{C adter wick}
\end{eqnarray}
where $ \Delta(P)=1/P^2 $, the trace ``$\, \underset{I,a}{\text{tr}} \,$'' is taken over the bispinor indices and the bispinor numbers $ I_n $, and the notation is introduced
\begin{eqnarray}
\langle T_{\tau} \Psi_{aI_1}(X)\overline{\Psi}_{bI_2}(Y)\rangle=\SumInt_{P} e^{i P (X-Y)} G_{ab (I_1 I_2)}(P)\,.
\label{prop four}
\end{eqnarray}
The summation over the Matsubara frequencies is performed according to the formulas \cite{Buzzegoli:2020fjm}
\begin{eqnarray}
\sum_{\omega_n = (2n+1)\pi T}\frac{f(\omega_n) e^{i \omega_n \tau}}{\omega_n^2+E^2}&=&\frac{1}{2E T}\sum_{s=\pm 1}f(-isE) e^{\tau s E} \Big[\theta(-s\tau)-n_F(E)\Big]\,,\nonumber \\
\sum_{\omega_n = (2n+1)\pi T}\frac{f(\omega_n) e^{i \omega_n \tau}}{(\omega_n^2+E^2)^2} &=&
\frac{1}{T}\sum_{s=\pm 1}e^{\tau s E}  \Bigg\{ \frac{f(-isE)}{4 E^2}n_F'(E)
+\frac{(1-s \tau E) f(-i s E) + i s E f'(-i s E)}{4 E^3}
 \Big[\theta(-s\tau)-n_F(E)\Big] \Bigg\}\,,
\label{mats sum}
\end{eqnarray}
where $ f(x) $ is an analytic function and $ n_F(E) ={(1+e^{E/T})^{-1}}$ is the Fermi-Dirac distribution. Each of the propagators in (\ref{C adter wick}) contains poles of either the 2nd or the 4th order
\begin{eqnarray}
G(P)=G_1(P)+G_2(P)\,,\,\, G_1\sim 1/P^2\,,\,\,G_2\sim 1/P^4\,.
\label{prop pol}
\end{eqnarray}
The summation in the contribution from only $ G_1(P) $ to (\ref{C split}) can be performed using the first of the formulas in (\ref{mats sum}), while the rest of the terms contain higher order poles, where the second formula needs to be used. Considering the contribution from $ G_1(P) $ only, we obtain
\begin{eqnarray}
C_{\text{Wick1,G1}}^{\alpha_1\alpha_2|\alpha_3\alpha_4|\alpha_5\alpha_6|\lambda|ijk} &=&-T^3 \int [d\tau]d^3 x d^3 y d^3 z\, x^i y^j z^k
\times \int \frac{d^3 p_1 d^3 p_2 d^3 p_3 d^3 p_4}{16 E_1 E_2 E_3 E_4} e^{-i \bold{p}_1(\bold{y}-\bold{x})-i \bold{p}_2\bold{x}-i \bold{p}_3(\bold{z}-\bold{y})+i \bold{p}_4\bold{z}}
 \nonumber \\ && \times
\sum_{ s_n=\pm 1 } e^{s_1 E_1(\tau_y-\tau_x) + s_2 E_2 \tau_x + s_3 E_3(\tau_z-\tau_y)- s_4 E_4 \tau_z}\,
\bigg\{\theta(-s_1[\tau_y-\tau_x])-n_F(E_1)\bigg\}\nonumber \\ && \times \,
 \bigg\{\theta(-s_2\tau_x)-n_F(E_2)\bigg\}\, 
 \bigg\{\theta(-s_3[\tau_z-\tau_y ]) - n_F(E_3)\bigg\}\,   \bigg\{\theta(-s_4\tau_z)-n_F(E_4)\bigg\}
 \nonumber \\ && \times\, {\underset{I,a}{\text{tr}}} \Bigg\{ \mathcal{D}^{\alpha_1\alpha_2}(-i \widetilde{P}_1, i \widetilde{P}_2) G_1(\widetilde{P}_2) \mathcal{J}^{\lambda} G_1(\widetilde{P}_4) \mathcal{D}^{\alpha_5\alpha_6}(-i \widetilde{P}_4, i \widetilde{P}_3)  G_1(\widetilde{P}_3)\mathcal{D}^{\alpha_3\alpha_4}(-i \widetilde{P}_3, i \widetilde{P}_1) G_1(\widetilde{P}_1)\Bigg\} \,,\qquad
\label{C after sum mats}
\end{eqnarray}
where $ \widetilde{P}^{n}_{\mu} = {(- i s_{n} E_{1},-\bold{p}_1)} $, $ E_n={|\bold{p}_n|} $. The coordinate dependence in (\ref{C after sum mats}) can be absorbed into momentum derivatives
\begin{eqnarray}
&&\int d^3 p_1 d^3 p_2 d^3 p_3 d^3 p_4 d^3 x d^3 y d^3 z\,  x^i y^j z^k
 e^{-i \bold{p}_1(\bold{y}-\bold{x})-i \bold{p}_2\bold{x}-i \bold{p}_3(\bold{z}-\bold{y})+i \bold{p}_4\bold{z}} f(\bold{p}_1, \bold{p}_2, \bold{p}_3, \bold{p}_4)=
\nonumber \\ && = i(2\pi)^9 \int d^3p \left(\frac{\partial^3}{\partial p_4^{k}\partial p_2^{i}\partial p_3^{j}}+\frac{\partial^3}{\partial p_4^{k}\partial p_2^{i}\partial p_4^{j}}\right)
f(\bold{p}_1,\bold{p}_2,\bold{p}_3,\bold{p}_4)\Big|\def\arraystretch{0.5}\begin{array}{ll}
{\scriptstyle \bold{p}_4=\bold{p}_1}\\
{\scriptstyle \bold{p}_3=\bold{p}_1}\\
{\scriptstyle \bold{p}_2=\bold{p}_1}
\end{array}\,.
\label{int parts}
\end{eqnarray}

Thus, to calculate (\ref{C adter wick}), it is necessary first to sum over the Matsubara frequencies using (\ref{mats sum}) and then use (\ref{int parts}). After that, it is necessary to calculate the trace of the product of the matrices ``$\, \underset{ I,a }{\text{tr}} \,$''. Finally, integration over $ \tau_{x}, \tau_{y} $ and $ \tau_{z} $ can be made explicitly, as well as integration over angles in $ d^3 p = \sin(\vartheta) p^2  dp \,d{\vartheta} \,d{\phi} $. At the end, we obtain the following integral, for example, for $ C^{02|02|02|3|111} $, which can be found analytically
\begin{eqnarray}
C^{02|02|02|3|111} &=& \frac{T}{480 \pi^2} \int \frac{dp\, p\, e^{p/T}}{(1+e^{p/T})^5}\Bigg\{ 
126-291\frac{p}{T}
-472 \frac{p^2}{T^2}
+
\bigg[
126+873\frac{p}{T}  +5192 \frac{p^2}{T^2}\bigg] e^{p/T}
\nonumber \\ &&+
\bigg[
-126+873\frac{p}{T}  -5192 \frac{p^2}{T^2}\bigg] e^{2p/T}
+
\bigg[
-126-291\frac{p}{T}
+472 \frac{p^2}{T^2}\bigg]e^{3p/T}
\Bigg\} = \frac{177 T^3}{80 \pi^2} \,.
\label{C res}
\end{eqnarray}
In this way all the coefficients $ C $ in (\ref{lambda C}) can be calculated. As a result, we will have
\begin{eqnarray}
\lambda_1 &=&-\frac{1}{6}\Bigg(2\cdot\frac{177}{80 \pi^2} +6 \cdot \frac{353}{240 \pi^2}\Bigg)=-\frac{53}{24 \pi^2} \,,
\nonumber \\
\lambda_2 &=&-\frac{1}{6}\Bigg(\frac{33}{40 \pi^2}
+\frac{53}{80 \pi^2}+\frac{1}{2 \pi^2}+
\frac{3}{4 \pi^2} +\frac{47}{80 \pi^2}+
\frac{17}{40 \pi^2}
\Bigg)
= -\frac{5}{8 \pi^2}\,.
\label{final sum beat}
\end{eqnarray}
Returning to (\ref{kve gen}), we obtain the KVE current in the RSA theory
\begin{eqnarray}
j^{\mu}_{A,KVE} = \left(-\frac{53}{24 \pi^2} \omega^2-\frac{5}{8 \pi^2} a^2\right)
\omega^{\mu}\,.
\label{kve 32}
\end{eqnarray}
Comparing (\ref{kve 32}) with (\ref{anom rsa}), we obtain
\begin{eqnarray}
\lambda_1-\lambda_2= -\frac{53}{24 \pi^2}+\frac{5}{8 \pi^2} =
-\frac{32 \cdot 19}{384 \pi^2}\,,
\label{check 32}
\end{eqnarray}
which confirms the general relation (\ref{theorem2}). Factor ``$ -19 $'' in (\ref{check 32}) is a clear indicator of the gravitational anomaly (\ref{anom rsa}). 

\section{Discussion}
\label{sec disc}

Let us consider the formula (\ref{kve 32}) in more detail. It was obtained by us in the $ \mu=0 $ approximation and in the third order in gradients. However, it was noticed that for the massless limit, many quantities are expressed by finite polynomials with positive powers of $ \omega, a, \mu, T ... $ \cite{Prokhorov:2019hif}. In particular, a similar formula for the Dirac field (\ref{kve 12}) was ``exact'' in the massless limit\footnote{(\ref{kve 12}) and (\ref{RSA exact}) are valid at a sufficiently high temperature \cite{Vilenkin:1979ui}. It was shown, that at $ T \sim |a|, |\omega| $ the system becomes unstable \cite{Prokhorov:2019hif}.}. We might assume, that (\ref{kve 32}) and (\ref{cve 32}) also lead to the ``all-orders'' expression
\begin{eqnarray}
j^{\mu}_{A,RSA} &=&\left(\frac{5}{6}T^2+\frac{5}{2\pi^2}\mu^2-\frac{53}{24 \pi^2}\omega^2 -\frac{5}{8 \pi^2}a^2 \right)
\omega^{\mu}\,.
\label{RSA exact}
\end{eqnarray}

The derivation described in Section \ref{sec gen deriv} is quite general and should be valid for an arbitrary spin $ S $. The anomaly for fields with an arbitrary spin is known from the supergravity \cite{Duff:1982yw, Christensen:1978md}
\begin{eqnarray}
\nabla_{\mu} j^{\mu}_{A,S} =
\frac{(S-2S^3)}{96 \pi^2\sqrt{-g}}\varepsilon^{\mu\nu\rho\sigma}
R_{\mu\nu\kappa\lambda}{R_{\rho\sigma}}^{\kappa\lambda}\,.
\label{anom grav gen S}
\end{eqnarray}
According to (\ref{theorem2}), the anomaly allows us to find only the combination of the currents $ \varpi_{\alpha\beta}\varpi^{\alpha\beta} \omega_{\mu} = \frac{2}{T^2}(a^2-\omega^2)\omega_{\mu} $, but not the conductivities of each of them separately. On the other hand, if we compare (\ref{kve 12}) and (\ref{kve 32}), then we can assume that $ \lambda_{2} \sim S $. Then immediately from (\ref{anom grav gen S}) and (\ref{theorem2}), we will have
\begin{eqnarray}
\lambda^S_{1} = \frac{S-8 S^3}{12 \pi^2}\,,\quad
\lambda^S_{2} = -\frac{S}{4 \pi^2}\,.
\label{hypo}
\end{eqnarray}
Hypothetical formulas (\ref{hypo}) satisfy general equations (\ref{anom grav gen S}) and (\ref{theorem2}), and special cases (\ref{kve 12}) and (\ref{kve 32})\footnote{Considering that there are two additional degrees of freedom with spin 1/2 in the RSA theory \cite{Prokhorov:2022rna}.}.

Note that $ \lambda_{1}(S) $ differs from a cubic dependence on spin, which could be naively expected bearing in mind the spin-dependent contribution of the angular velocity to the Hamiltonian $ \Delta H =- \bold{\Omega} \cdot \bold{S} $ and the known nonperturbative formulas for spin 1/2 (see, e.g., \cite{Vilenkin:1980zv, Prokhorov:2018bql}). 

Using (\ref{theorem2}), (\ref{lambda C}) and (\ref{C}) one can find the gravitational chiral anomaly for various field theories (that do not have dimensional parameters) by calculating four-point functions in ordinary flat space-time and finite temperature. This approach can be compared with the one described in \cite{Erdmenger:1999xx}, according to which the three-point correlator have the form of a product of a universal function and a numerical factor equal to the coefficient in the gravitational anomaly.


\section{Conclusion}
\label{sec concl}

In this paper, we have considered relativistic hydrodynamics for particles with spins 3/2 and 1/2, described by the Rarita-Schwinger-Adler field theory. We have shown that the relationship between the hydrodynamic transport coefficients, which follows from gravity and the gravitational chiral anomaly, is confirmed. For this, the coefficients with rotation and acceleration in the third order of the gradient expansion, corresponding to the KVE, were found using the Zubarev density operator. It is demonstrated that that they reproduce the factor ``$ -19 $'' from the gravitational chiral anomaly, confirming the general formula. Thus, the cubic dependence on spin, which leads to a rapid growth of the anomalous factor, is clearly visible in hydrodynamics.

Using the known result for the gravitational anomaly for an arbitrary spin, an interpolating formulas are constructed for the cubic transport coefficients as spin functions. 

The relationship between the gravitational anomaly and KVE opens up an opportunity to search for the gravitational anomaly manifestations outside the curved space-time, in systems with extremely large acceleration and vorticity. Also it can be used to find the gravitational chiral anomaly from the flat-space four-point functions.

{\bf Acknowledgements}

The authors are thankful to M. J. Duff for valuable comments. The work was supported by Russian Science Foundation Grant No. 21-12-00237. 

\bibliography{lit}

\begin{thebibliography}{10}
\expandafter\ifx\csname url\endcsname\relax
  \def\url#1{\texttt{#1}}\fi
\expandafter\ifx\csname urlprefix\endcsname\relax\def\urlprefix{URL }\fi
\expandafter\ifx\csname href\endcsname\relax
  \def\href#1#2{#2} \def\path#1{#1}\fi

\bibitem{Son:2009tf}
D.~T. Son, P.~Surowka, {Hydrodynamics with Triangle Anomalies}, Phys. Rev.
  Lett. 103 (2009) 191601.
\newblock \href {http://arxiv.org/abs/0906.5044} {\path{arXiv:0906.5044}},
  \href {https://doi.org/10.1103/PhysRevLett.103.191601}
  {\path{doi:10.1103/PhysRevLett.103.191601}}.

\bibitem{Fukushima:2008xe}
K.~Fukushima, D.~E. Kharzeev, H.~J. Warringa, {The Chiral Magnetic Effect},
  Phys. Rev. D78 (2008) 074033.
\newblock \href {http://arxiv.org/abs/0808.3382} {\path{arXiv:0808.3382}},
  \href {https://doi.org/10.1103/PhysRevD.78.074033}
  {\path{doi:10.1103/PhysRevD.78.074033}}.

\bibitem{Sadofyev:2010is}
A.~V. Sadofyev, V.~I. Shevchenko, V.~I. Zakharov, {Notes on chiral
  hydrodynamics within effective theory approach}, Phys. Rev. D83 (2011)
  105025.
\newblock \href {http://arxiv.org/abs/1012.1958} {\path{arXiv:1012.1958}},
  \href {https://doi.org/10.1103/PhysRevD.83.105025}
  {\path{doi:10.1103/PhysRevD.83.105025}}.

\bibitem{Buzzegoli:2020ycf}
M.~Buzzegoli, {Thermodynamic equilibrium of massless fermions with vorticity,
  chirality and electromagnetic field}, Lect. Notes Phys. 987 (2021) 53--93.
\newblock \href {http://arxiv.org/abs/2011.09974} {\path{arXiv:2011.09974}},
  \href {https://doi.org/10.1007/978-3-030-71427-7_3}
  {\path{doi:10.1007/978-3-030-71427-7_3}}.

\bibitem{Yang:2022ksq}
S.-Z. Yang, J.-H. Gao, Z.-T. Liang, {Constraining Non-Dissipative Transport
  Coefficients in Global Equilibrium}, Symmetry 14~(5) (2022) 948.
\newblock \href {http://arxiv.org/abs/2203.14023} {\path{arXiv:2203.14023}},
  \href {https://doi.org/10.3390/sym14050948} {\path{doi:10.3390/sym14050948}}.

\bibitem{Mitra:2021wjp}
A.~K. Mitra, S.~Ghosh, {Divergence anomaly and Schwinger terms: Towards a
  consistent theory of anomalous classical fluids}, Phys. Rev. D 106~(4) (2022)
  L041702.
\newblock \href {http://arxiv.org/abs/2111.00473} {\path{arXiv:2111.00473}},
  \href {https://doi.org/10.1103/PhysRevD.106.L041702}
  {\path{doi:10.1103/PhysRevD.106.L041702}}.

\bibitem{Nair:2011mk}
V.~P. Nair, R.~Ray, S.~Roy, {Fluids, Anomalies and the Chiral Magnetic Effect:
  A Group-Theoretic Formulation}, Phys. Rev. D 86 (2012) 025012.
\newblock \href {http://arxiv.org/abs/1112.4022} {\path{arXiv:1112.4022}},
  \href {https://doi.org/10.1103/PhysRevD.86.025012}
  {\path{doi:10.1103/PhysRevD.86.025012}}.

\bibitem{Nissinen:2021gke}
J.~Nissinen, G.~E. Volovik, {Anomalous chiral transport with vorticity and
  torsion: Cancellation of two mixed gravitational anomaly currents in rotating
  chiral p+ip Weyl condensates}, Phys. Rev. D 106~(4) (2022) 045022.
\newblock \href {http://arxiv.org/abs/2111.08639} {\path{arXiv:2111.08639}},
  \href {https://doi.org/10.1103/PhysRevD.106.045022}
  {\path{doi:10.1103/PhysRevD.106.045022}}.

\bibitem{Chernodub:2021nff}
M.~N. Chernodub, Y.~Ferreiros, A.~G. Grushin, K.~Landsteiner, M.~A.~H.
  Vozmediano, {Thermal transport, geometry, and anomalies} (10 2021).
\newblock \href {http://arxiv.org/abs/2110.05471} {\path{arXiv:2110.05471}}.

\bibitem{Rogachevsky:2010ys}
O.~Rogachevsky, A.~Sorin, O.~Teryaev, {Chiral vortaic effect and neutron
  asymmetries in heavy-ion collisions}, Phys. Rev. C82 (2010) 054910.
\newblock \href {http://arxiv.org/abs/1006.1331} {\path{arXiv:1006.1331}},
  \href {https://doi.org/10.1103/PhysRevC.82.054910}
  {\path{doi:10.1103/PhysRevC.82.054910}}.

\bibitem{Becattini:2022zvf}
F.~Becattini, {Spin and polarization: a new direction in relativistic heavy ion
  physics} (4 2022).
\newblock \href {http://arxiv.org/abs/2204.01144} {\path{arXiv:2204.01144}}.

\bibitem{STAR:2017ckg}
L.~Adamczyk, et~al., {Global $\Lambda$ hyperon polarization in nuclear
  collisions: evidence for the most vortical fluid}, Nature 548 (2017) 62--65.
\newblock \href {http://arxiv.org/abs/1701.06657} {\path{arXiv:1701.06657}},
  \href {https://doi.org/10.1038/nature23004} {\path{doi:10.1038/nature23004}}.

\bibitem{STAR:2020xbm}
J.~Adam, et~al., {Global Polarization of $\Xi$ and $\Omega$ Hyperons in Au+Au
  Collisions at $\sqrt {s_{NN}}$ = 200 GeV}, Phys. Rev. Lett. 126~(16) (2021)
  162301.
\newblock \href {http://arxiv.org/abs/2012.13601} {\path{arXiv:2012.13601}},
  \href {https://doi.org/10.1103/PhysRevLett.126.162301}
  {\path{doi:10.1103/PhysRevLett.126.162301}}.

\bibitem{Landau1987Fluid}
L.~D. Landau, E.~M. Lifshitz, Fluid Mechanics, 2nd Edition, Vol.~6 of Course of
  theoretical physics, Butterworth-Heinemann, 1987.

\bibitem{Alvarez-Gaume:1984}
L.~Alvarez-Gaume, E.~Witten, {Gravitational Anomalies}, Nucl. Phys. B234 (1984)
  269, [269(1983)].
\newblock \href {https://doi.org/10.1016/0550-3213(84)90066-X}
  {\path{doi:10.1016/0550-3213(84)90066-X}}.

\bibitem{Duff:1982yw}
M.~J. Duff, {Ultraviolet divergences in extended supergravity}, in: {First
  School on Supergravity}, 1982.
\newblock \href {http://arxiv.org/abs/1201.0386} {\path{arXiv:1201.0386}}.

\bibitem{Landsteiner:2011cp}
K.~Landsteiner, E.~Megias, F.~Pena-Benitez, {Gravitational Anomaly and
  Transport}, Phys. Rev. Lett. 107 (2011) 021601.
\newblock \href {http://arxiv.org/abs/1103.5006} {\path{arXiv:1103.5006}},
  \href {https://doi.org/10.1103/PhysRevLett.107.021601}
  {\path{doi:10.1103/PhysRevLett.107.021601}}.

\bibitem{Jensen:2012kj}
K.~Jensen, R.~Loganayagam, A.~Yarom, {Thermodynamics, gravitational anomalies
  and cones}, JHEP 02 (2013) 088.
\newblock \href {http://arxiv.org/abs/1207.5824} {\path{arXiv:1207.5824}},
  \href {https://doi.org/10.1007/JHEP02(2013)088}
  {\path{doi:10.1007/JHEP02(2013)088}}.

\bibitem{Stone:2018zel}
M.~Stone, J.~Kim, {Mixed Anomalies: Chiral Vortical Effect and the Sommerfeld
  Expansion}, Phys. Rev. D98~(2) (2018) 025012.
\newblock \href {http://arxiv.org/abs/1804.08668} {\path{arXiv:1804.08668}},
  \href {https://doi.org/10.1103/PhysRevD.98.025012}
  {\path{doi:10.1103/PhysRevD.98.025012}}.

\bibitem{Prokhorov:2020npf}
G.~Y. Prokhorov, O.~V. Teryaev, V.~I. Zakharov, {Chiral vortical effect for
  vector fields}, Phys. Rev. D 103~(8) (2021) 085003.
\newblock \href {http://arxiv.org/abs/2009.11402} {\path{arXiv:2009.11402}},
  \href {https://doi.org/10.1103/PhysRevD.103.085003}
  {\path{doi:10.1103/PhysRevD.103.085003}}.

\bibitem{Adler:2017shl}
S.~L. Adler, {Analysis of a gauged model with a spin-$\frac{1}{2}$ field
  directly coupled to a Rarita-Schwinger spin-$\frac{3}{2}$ field}, Phys. Rev.
  D 97~(4) (2018) 045014.
\newblock \href {http://arxiv.org/abs/1711.00907} {\path{arXiv:1711.00907}},
  \href {https://doi.org/10.1103/PhysRevD.97.045014}
  {\path{doi:10.1103/PhysRevD.97.045014}}.

\bibitem{Adler:2019zxx}
S.~L. Adler, P.~Pais, {Chiral anomaly calculation in the extended coupled
  Rarita-Schwinger model}, Phys. Rev. D 99~(9) (2019) 095037.
\newblock \href {http://arxiv.org/abs/1903.06189} {\path{arXiv:1903.06189}},
  \href {https://doi.org/10.1103/PhysRevD.99.095037}
  {\path{doi:10.1103/PhysRevD.99.095037}}.

\bibitem{Prokhorov:2022rna}
G.~Y. Prokhorov, O.~V. Teryaev, V.~I. Zakharov, {Gravitational chiral anomaly
  for spin 3/2 field interacting with spin 1/2 field}, Phys. Rev. D 106~(2)
  (2022) 025022.
\newblock \href {http://arxiv.org/abs/2202.02168} {\path{arXiv:2202.02168}},
  \href {https://doi.org/10.1103/PhysRevD.106.025022}
  {\path{doi:10.1103/PhysRevD.106.025022}}.

\bibitem{Prokhorov:2021bbv}
G.~Y. Prokhorov, O.~V. Teryaev, V.~I. Zakharov, {Chiral vortical effect in
  extended Rarita-Schwinger field theory and chiral anomaly}, Phys. Rev. D
  105~(4) (2022) L041701.
\newblock \href {http://arxiv.org/abs/2109.06048} {\path{arXiv:2109.06048}},
  \href {https://doi.org/10.1103/PhysRevD.105.L041701}
  {\path{doi:10.1103/PhysRevD.105.L041701}}.

\bibitem{Prokhorov:2022udo}
G.~Y. Prokhorov, O.~V. Teryaev, V.~I. Zakharov, {Hydrodynamic Manifestations of
  Gravitational Chiral Anomaly}, Phys. Rev. Lett. 129~(15) (2022) 151601.
\newblock \href {http://arxiv.org/abs/2207.04449} {\path{arXiv:2207.04449}},
  \href {https://doi.org/10.1103/PhysRevLett.129.151601}
  {\path{doi:10.1103/PhysRevLett.129.151601}}.

\bibitem{Becattini:2014yxa}
F.~Becattini, L.~Bucciantini, E.~Grossi, L.~Tinti, {Local thermodynamical
  equilibrium and the beta frame for a quantum relativistic fluid}, Eur. Phys.
  J. C 75~(5) (2015) 191.
\newblock \href {http://arxiv.org/abs/1403.6265} {\path{arXiv:1403.6265}},
  \href {https://doi.org/10.1140/epjc/s10052-015-3384-y}
  {\path{doi:10.1140/epjc/s10052-015-3384-y}}.

\bibitem{Prokhorov:2018bql}
G.~Y. Prokhorov, O.~V. Teryaev, V.~I. Zakharov, {Effects of rotation and
  acceleration in the axial current: density operator vs Wigner function}, JHEP
  02 (2019) 146.
\newblock \href {http://arxiv.org/abs/1807.03584} {\path{arXiv:1807.03584}},
  \href {https://doi.org/10.1007/JHEP02(2019)146}
  {\path{doi:10.1007/JHEP02(2019)146}}.

\bibitem{Ambrus:2021eod}
V.~E. Ambrus, E.~Winstanley, {Vortical Effects for Free Fermions on Anti-De
  Sitter Space-Time}, Symmetry 13 (2021) 2019.
\newblock \href {http://arxiv.org/abs/2107.06928} {\path{arXiv:2107.06928}},
  \href {https://doi.org/10.3390/sym13112019} {\path{doi:10.3390/sym13112019}}.

\bibitem{Ambrus:2019ayb}
V.~E. Ambrus, {Helical massive fermions under rotation}, JHEP 08 (2020) 016.
\newblock \href {http://arxiv.org/abs/1912.09977} {\path{arXiv:1912.09977}},
  \href {https://doi.org/10.1007/JHEP08(2020)016}
  {\path{doi:10.1007/JHEP08(2020)016}}.

\bibitem{Palermo:2021hlf}
A.~Palermo, M.~Buzzegoli, F.~Becattini, {Exact equilibrium distributions in
  statistical quantum field theory with rotation and acceleration: Dirac
  field}, JHEP 10 (2021) 077.
\newblock \href {http://arxiv.org/abs/2106.08340} {\path{arXiv:2106.08340}},
  \href {https://doi.org/10.1007/JHEP10(2021)077}
  {\path{doi:10.1007/JHEP10(2021)077}}.

\bibitem{Vilenkin:1979ui}
A.~Vilenkin, {Macroscopic parity violating effects: neutrino fluxes from
  rotating black holes and in rotating thermal radiation}, Phys. Rev. D20
  (1979) 1807--1812.
\newblock \href {https://doi.org/10.1103/PhysRevD.20.1807}
  {\path{doi:10.1103/PhysRevD.20.1807}}.

\bibitem{Vilenkin:1980zv}
A.~Vilenkin, {Quantum field theory at finite temperature in a rotating system},
  Phys. Rev. D21 (1980) 2260--2269.
\newblock \href {https://doi.org/10.1103/PhysRevD.21.2260}
  {\path{doi:10.1103/PhysRevD.21.2260}}.

\bibitem{Buzzegoli:2017cqy}
M.~Buzzegoli, E.~Grossi, F.~Becattini, {General equilibrium second-order
  hydrodynamic coefficients for free quantum fields}, JHEP 10 (2017) 091,
  [Erratum: JHEP07,119(2018)].
\newblock \href {http://arxiv.org/abs/1704.02808} {\path{arXiv:1704.02808}},
  \href {https://doi.org/10.1007/JHEP07(2018)119, 10.1007/JHEP10(2017)091}
  {\path{doi:10.1007/JHEP07(2018)119, 10.1007/JHEP10(2017)091}}.

\bibitem{Zubarev:1979}
D.~N. Zubarev, A.~V. Prozorkevich, S.~S. A., {Derivation of nonlinear
  generalized equations of quantum relativistic hydrodynamics}, TMF 40:3 (1979)
  394–407.

\bibitem{Prokhorov:2019cik}
G.~Y. Prokhorov, O.~V. Teryaev, V.~I. Zakharov, {Unruh effect for fermions from
  the Zubarev density operator}, Phys. Rev. D99~(7) (2019) 071901(R).
\newblock \href {http://arxiv.org/abs/1903.09697} {\path{arXiv:1903.09697}},
  \href {https://doi.org/10.1103/PhysRevD.99.071901}
  {\path{doi:10.1103/PhysRevD.99.071901}}.

\bibitem{Prokhorov:2019yft}
G.~Y. Prokhorov, O.~V. Teryaev, V.~I. Zakharov, {Unruh effect universality:
  emergent conical geometry from density operator}, JHEP 03 (2020) 137.
\newblock \href {http://arxiv.org/abs/1911.04545} {\path{arXiv:1911.04545}},
  \href {https://doi.org/10.1007/JHEP03(2020)137}
  {\path{doi:10.1007/JHEP03(2020)137}}.

\bibitem{Laine:2016hma}
M.~Laine, A.~Vuorinen, {Basics of Thermal Field Theory}, Lect. Notes Phys. 925
  (2016) pp.1--281.
\newblock \href {http://arxiv.org/abs/1701.01554} {\path{arXiv:1701.01554}},
  \href {https://doi.org/10.1007/978-3-319-31933-9}
  {\path{doi:10.1007/978-3-319-31933-9}}.

\bibitem{Buzzegoli:2020fjm}
M.~Buzzegoli, {Thermodynamic equilibrium of massless fermions with vorticity,
  chirality and magnetic field}, Other thesis (4 2020).
\newblock \href {http://arxiv.org/abs/2004.08186} {\path{arXiv:2004.08186}}.

\bibitem{Prokhorov:2019hif}
G.~Y. Prokhorov, O.~V. Teryaev, V.~I. Zakharov, {Thermodynamics of accelerated
  fermion gases and their instability at the Unruh temperature}, Phys. Rev.
  D100~(12) (2019) 125009.
\newblock \href {http://arxiv.org/abs/1906.03529} {\path{arXiv:1906.03529}},
  \href {https://doi.org/10.1103/PhysRevD.100.125009}
  {\path{doi:10.1103/PhysRevD.100.125009}}.

\bibitem{Christensen:1978md}
S.~M. Christensen, M.~J. Duff, {New Gravitational Index Theorems and
  Supertheorems}, Nucl. Phys. B 154 (1979) 301--342.
\newblock \href {https://doi.org/10.1016/0550-3213(79)90516-9}
  {\path{doi:10.1016/0550-3213(79)90516-9}}.

\bibitem{Erdmenger:1999xx}
J.~Erdmenger, {Gravitational axial anomaly for four-dimensional conformal field
  theories}, Nucl. Phys. B 562 (1999) 315--329.
\newblock \href {http://arxiv.org/abs/hep-th/9905176}
  {\path{arXiv:hep-th/9905176}}, \href
  {https://doi.org/10.1016/S0550-3213(99)00561-1}
  {\path{doi:10.1016/S0550-3213(99)00561-1}}.

\end{thebibliography}

\end{document}